\documentclass[pra,amsmath,a4paper,twocolumn]{revtex4}
\usepackage{graphicx}
\usepackage{amsfonts}
\usepackage{amsmath}
\usepackage{ltxtable}
\usepackage{CJK}
\usepackage{mdwmath}
\usepackage{color}

\newcommand{\mX}{{\mathcal X}}

\newcommand{\mM}{{\mathcal M}}

\begin{document}

\title{Multi-setting  Greenberger-Horne-Zeilinger paradoxes}
\author{ Weidong Tang$^{1,2}$, Sixia Yu$^{2,3}$ and C.H. Oh$^{3}$}
\affiliation{$^1$Key Laboratory of Quantum Information and Quantum Optoelectronic Devices, Shaanxi Province,
 and Department of Applied Physics
of Xi'an Jiaotong University, Xi'an 710049, P.R. China\\
$^2$Hefei National Laboratory for Physical Sciences at
Microscale and Department of Modern Physics
of University of Science and Technology of China, Hefei 230026, P.R. China\\
$^3$Centre for Quantum Technologies and Physics Department,
National University of Singapore, 2 Science Drive 3, Singapore 117542}

\begin{abstract}
Greenberger-Horne-Zeilinger (GHZ) paradox provides an all-versus-nothing test for the quantum nonlocality. In all the GHZ paradoxes known so far each observer is allowed to measure only two alternative observables. Here we shall present a general construction for GHZ paradoxes in which each observer measuring more than two observables given that the system is prepared in the $n$-qudit GHZ state. By doing so we are able to construct a multi-setting GHZ paradox for the $n$-qubit GHZ state, with $n$ being arbitrary, which is genuine $n$-partite, i.e., no GHZ paradox exists when restricted to a subset of number of observers for a given set of Mermin observables. Our result  fills up the gap of the absence of a genuine GHZ paradox for the GHZ state of an even number of qubits, especially the four-qubit GHZ state as used in GHZ's original proposal.

\end{abstract}

\maketitle

\section{Introduction}
Quantum nonlocality, or more generally contextuality, is a fascinating quantum feature that counters a comfortable classical intuition known as local realism introduced by Einstein, Podolsky, and Rosen's (EPR) via ``elements of reality" \cite{EPR} or more generally noncontextuality. Local realism or noncontextuality states that the result of a measurement cannot be affected by any spacelike separated events or which compatible observables might be measure alongside. This clashing between quantum mechanics and local realism can be  revealed by several ingenious approaches such as Bell's inequalities \cite{Bell} and Kochen-Specker (KS) theorem \cite{KS}, which have been verified in experiments on various physical systems \cite{bell exp, Duan12}. Compared to Bell's nonlocality, Greenberger-Horne-Zeilinger (GHZ) paradoxes \cite{GHZ,GHZ2}  reveal a stronger quantum nonlocality, known as GHZ nonlocality,  and provide an all-versus-nothing test of nonlocality \cite{first,GHZ exp1,GHZ exp2}.

Soon after its first discovery in a $4$-qubit GHZ state \cite{GHZ} the GHZ paradox is provided by Mermin  \cite{mermin} a compelling argument with the help of a set of mutually commuting observables, called Mermin observables,
\begin{align}\label{stab eq}
\mM= \{X_1Y_2Y_3,
 Y_1X_2Y_3,
 Y_1Y_2X_3,
 -X_1X_2X_3\}
\end{align}
that stabilize the $3$-qubit GHZ state $|\Phi\rangle=\frac{1}{\sqrt{2}}(|000\rangle-|111\rangle)$, i.e., $\mM_k|\Phi\rangle=|\Phi\rangle$, with
 $X_k, Y_k$ and $Z_k$ denoting  three pauli matrices  of the $k$-th qubit. Because of these perfect correlations the outcome of the measuring  any local observable $X_k$ or $Y_k$ can be predicted from the outcomes of the other two local observables (appears in corresponding $\mM_k$) in a space-like distance and thus, according to EPR's definition of elements of reality, both observables $X_k$ and $Y_k$  are elements of reality for all $k$ and must have realistic values
$m_i^x$ and $m_i^y$ for $i=1,2,3$. Moreover since all realistic values are supposed to obey the same algebraic constrains as their operators do all the following four products
\begin{align}
 m_1^xm_2^ym_3^y,\;
 m_1^ym_2^xm_3^y,\;
 m_1^ym_2^ym_3^x,\;
-m_1^xm_2^xm_3^x
 \end{align}
must assume value 1 because all Mermin observables stabilize the GHZ state. However this is impossible because  $(m_1^xm_2^ym_3^y)(m_1^ym_2^xm_3^y)(m_1^ym_2^ym_3^x)=m_1^xm_2^xm_3^x$, which leads to $-1=1$, a contradiction.

Besides its elegance and beauty, the GHZ paradox has found numerous applications such as in the quantum protocols for reducing communication complexity \cite{cleve} and for secret sharing \cite{secret}. Deeper understanding or a detailed classification of quantum nonlocality may be gained by searching for GHZ paradoxes in the case of multipartite and multilevel systems. Earlier generalizations of GHZ paradoxes to the case of multipartite and multilevel systems, e.g., \cite{cabello,pagonis},  can be reduced either to the qubit cases or to fewer particle cases. Genuine multipartite multilevel GHZ paradoxes had not been found  until a $(d+1)$-partite $d$-level systems with $d$ being even was raised by Cerf et al \cite{cerf}.
Lee et al proposed an unconventional approach by using concurrent observables, not commuting yet sharing a common eigenstate, to construct GHZ paradox for the GHZ states of an odd number of particles \cite{lee}.

There are  states other than GHZ states that can exhibit GHZ nonlocality. DiVincenzo and Peres \cite{divi} found that those special kind of entangled states used in a quantum error-correcting codes \cite{QECC}, which are equivalent to graph states, can also exhibit GHZ nonlocality. Most recently we have identified a family of weighted graphs, called as GHZ graphs, whose corresponding qudit graphs states may lead to genuine multipartite and multilevel GHZ paradoxes \cite{TYO}. Genuine GHZ paradoxes have been constructed for all even number of particles except for a few cases, e.g., qubits. As noticed in \cite{TYO} a genuine multipartite GHZ paradox for the 4-qubit GHZ state, which is used originally to demonstrate GHZ paradox, is ironically missing.

We note that all previous constructions of GHZ paradoxes involve only two settings, i.e., each observer measures only two alternative observables. And this observation turns out to be the key to the solution of above mentioned open problem. In this paper we shall at first present a general constructions of GHZ paradoxes for the qudit GHZ states  in which each observer measures more than two observables. And then we construct a genuine multipartite GHZ  paradoxes for the GHZ state of any number (larger than $2$) of qubits, even or odd. Lastly we derive a multisetting Bell inequality from the multi-setting GHZ paradoxes as an application.

\section{Genuine multi-setting GHZ paradoxes}

Consider a system of $n$ qudit with $d$ energy levels. Let  $I$ be the identity operator for a single qudit and define a standard unitary observable $\hat U$ to be a unitary qudit operator whose eigenvalues are integer powers of $\omega=e^{i2\pi/d}$, i.e., $\hat U^d=I$. For an arbitrary real number $r$ the following unitary observable
\begin{equation}
  \hat X_r=\omega^r\left(
e^{-i2\pi r}|d-1\rangle\langle 0|+\sum_{s=1}^{d-1}|s-1\rangle\langle s|\right)
\end{equation}
is standard. Let $\vec r=(r_1,r_2,\ldots,r_n)$ be an $n$-dimensional real vector and denote by $\vec 1$ and $\vec 0$ two $n$-dimensional vectors with all entries being equal to 1 and 0, respectively. The unitary observable
\begin{equation}
\hat X({\vec r}):=\bigotimes_{k=1}^n\hat X_{r_k}
\end{equation}
satisfies also $[\hat X({\vec r})]^d=I$ and, if  $R=\vec 1\cdot \vec r$ is an integer, has the $n$-qudit GHZ state $|\Phi\rangle$ as an eigenstate corresponding to eigenvalue $\omega^R$ with
\begin{equation}\label{GHZ state}
|\Phi\rangle=\frac{1}{\sqrt{d}}\sum_{i=0}^{d-1}|i\rangle_1\otimes|i\rangle_2\otimes\ldots\otimes|i\rangle_n.
\end{equation}
For convenience we simply denote by $\hat X=\hat X({\vec 0})$ the multiqudit shift operator.

To begin with let us see an example of a 3-setting GHZ paradox for an even number of qudits.
For an $n$-dimensional vector $\vec r$ we denote by $\sigma_k(\vec r)$ the vector obtained by cyclic (right) shifting $\vec r$ on $k$ positions with $k=0,1,\ldots, n-1$. That is the $j$-th component of $\sigma_k(\vec r)$ is equal to the $(j-k)$-th component of $\vec r$, i.e., $[\sigma_k(\vec r)]_j=[\vec r]_{j-k}$ with addition modulo $n$. Consider the following special $n$-dimensional vector
\begin{equation}
\vec r_0=\left(0,b,c,c,\ldots,c\right),\quad b=\frac{1-d}n,\quad c=\frac{d-1}{n(n-2)}
\end{equation}
and its $n-1$ cyclic permutations $\vec r_k=\sigma_k(\vec r)$ with $k=1,\ldots, n-1$. Because $R_k=\vec 1\cdot\vec r_k=0$ for $k=0,\ldots, n-1$,  all $n$ observables $\hat X({\vec r_k})$ stabilize the GHZ state $|\Phi\rangle$. Furthermore we have $\hat X(b\vec 1)|\Phi\rangle=\omega|\Phi\rangle$. We claim that a GHZ paradox arises from the following $n+2$ observables
\begin{equation}
\begin{array}{rc}
\hat X(\vec r_0)=&\hat X_0\otimes\hat  X_b\otimes\hat  X_c\otimes\hat  X_c\otimes\ldots\otimes\hat  X_c\otimes\hat  X_c\\
\hat X(\vec r_1)=&\hat X_c\otimes\hat  X_0\otimes\hat  X_b\otimes\hat  X_c\otimes\ldots\otimes\hat  X_c\otimes\hat  X_c\\
\hat X(\vec r_2)=&\hat  X_c\otimes\hat  X_c\otimes\hat X_0\otimes\hat  X_b\otimes\ldots\otimes\hat  X_c\otimes\hat  X_c\\
&\ldots\\
\hat X(\vec r_{n-2})=&\hat  X_c\otimes\hat  X_c\otimes\hat X_c\otimes\hat  X_c\otimes\ldots\otimes\hat  X_0\otimes\hat  X_b\\
\hat X(\vec r_{n-1})=&\hat  X_b\otimes\hat  X_c\otimes\hat X_c\otimes\hat  X_c\otimes\ldots\otimes\hat  X_c\otimes\hat  X_0\\
\hat X(\vec 0)=&\hat  X_0\otimes\hat  X_0\otimes\hat X_0\otimes\hat  X_0\otimes\ldots\otimes\hat  X_0\otimes\hat  X_0\cr
\hat X(b\vec 1)=&\hat  X_b\otimes\hat  X_b\otimes\hat X_b\otimes\hat  X_b\otimes\ldots\otimes\hat  X_b\otimes\hat  X_b
\end{array}.
\end{equation}

 For each qudit there are 3 different local observables $\hat X_{0,b,c}$ appearing in those $n+2$ observables given above. Let $m^{0,b,c}_j$ be the realistic local value for the observable $\hat X_{0,b,c}$ of the $j$-th qudit, which is some integer power of $\omega$. It is easy to see that both observables $\hat X_0, \hat X_b$ appear twice while observable $\hat X_c$ appears $n-2$ times which is also even since the number of qudits is even. From the quantum mechanical identities $\hat X|\Phi\rangle=|\Phi\rangle, \hat X({b\vec 1})|\Phi\rangle=\omega|\Phi\rangle$, and $\hat X(\vec r_k)|\Phi\rangle=|\Phi\rangle$ it follows immediately the following constraint
\begin{equation}
\prod_{j=1}^n\left(m_j^0m_j^b\right)^2\left(m_j^c\right)^{n-2}=\omega
\end{equation}
on local realistic values. This equation is impossible  when $d$ is even
because its l.h.s is an even power of $\omega$ while its r.h.s is an odd power of $\omega$. Or more explicitly, the $d/2$ power to the both sides of the above equality yields a contradiction $1=-1$.

The generalization of above construction of GHZ paradox to more than 3 settings is straightforward. For an $n$-dimensional vector $\vec r=(r_1,r_2,\ldots, r_n)$ we denote by $Y_{\vec r}:=\{y_1,y_2,\ldots,y_t\}$ the set of different entries appearing in $\vec r$, i.e., $y_k\not =y_j$ if $k\not=j$ and  for every $k$ there is a $y\in Y_{\vec r}$ such that $r_k=y$. For each $y\in Y_{\vec r}$ we denote by $g_y$ the number of times that the entry $y$ appears in $\vec r$.
 A vector $\vec r$ is defined to be a {\it GHZ vector} if {\it i) $b=(1-d)/n\in Y_{\vec r}$ and appears in $\vec r$ an odd number of times, i.e., $g_b$ is odd; ii) any other nonzero $y\in Y_{\vec r}$ appears in $\vec r$ an even number of times, i.e., $g_y$ is even (at least 2) if $y\not=b,0$; iii) $\vec 1\cdot \vec r=0$.}

For any GHZ vector $\vec r$ we denote $\vec r_0=\vec r$ and $\vec r_k=\sigma_k(\vec r)$ the $k$-shift vector of $\vec r$ for $k=1,2,\ldots,n-1$. All $n$ observables $\hat X({\vec r_k})$ stabilize the GHZ state due to requirement iii). We claim that the following set of $n+\sigma$ observables, with $\sigma=1,2 $ when $n$ is odd or even, respectively,
\begin{equation}\label{x}
\mX=\{\hat X({\vec r_0}),\hat X({\vec r_1}),\ldots,\hat X({\vec r_{n-1}}),\hat X^{\sigma-1},\hat X({b\vec 1})\}
\end{equation}
provides a GHZ paradox with $t=|Y_{\vec r}|$ settings.
We note at first that for each qudit there are exactly $t$ different local observables $\hat X_y$ with $y\in Y_{\vec r}$ and each one of those $t$ observables appears an even number of times in $n+\sigma$ observables in $\mX$. In fact for a nonzero $y\not=b\in Y_{\vec r}$ the observable $\hat X_y$ appears an even number of times by the definition of a GHZ vector (requirement ii)). By requirement i) the observable $\hat X_b$ for each qudit appears in the collection of $n$ observables $\hat X_{\vec r_k}$ an odd number of times and once in $\hat X_{b\vec 1}$, also an even number of times. Observable $\hat X_0$ appears in $\hat X_{\vec r_k}$ an even (odd) number of times if $n$ is odd (even), i.e., $g_0+\sigma-1$ is always even due to the additional observable $\hat X$ in the case of even $n$. Let $m_j^y$ be the local realistic values for the observable $\hat X_y$ for each $y\in Y_{\vec r}$. From the facts that $\hat X$ and $\hat X({\vec r_k})$ stabilize the GHZ state $|\Phi\rangle$ and $\hat X({b\vec 1})|\Phi\rangle=\omega|\Phi\rangle$ it follows the following constraint
\begin{equation}\prod_{j=1}^n(m_j^b)^{g_b+1}(m_j^0)^{g_0+\sigma-1}\prod_{y\in Y_{\vec r},y\not=b,0}\left(m_j^y\right)^{g_y}=\omega,\end{equation}
which is impossible for even $d$. Thus we have

{\it Theorem $1$ } Let $d$ be even and  $\sigma=1,2$ if $n$ is odd or even, respectively. Let
 $\vec r=\vec r_0$ be an arbitrary  $n$-dimensional GHZ vector  with $t$ different entries and $\vec r_k$ be its cyclic shifted vector by $k$ positions.  The set $\mX$ of $n+\sigma$  observables as given by Eq.(\ref{x})  provides a $t$-setting GHZ paradox for $n$-qudit GHZ state.

The issue of genuine multipartite and multilevel GHZ paradox is raised by Cerf etal \cite{cerf}.  A GHZ paradox is said to be genuine $n$-partite if one cannot reduce the
parties and still has a GHZ paradox. It should be noted that there are two necessary conditions for a set of observables to provide a GHZ argument. First, no local observable appears only once. Second these observables, commuting or not, should share a common eigenstate.

{\it Theorem $2$ } Let $s=\left\lfloor \frac {n-1}2 \right\rfloor $ and  $n=2s+\sigma$ with $\sigma=1,2$ if $n$ is odd or even.  The $n$-dimensional real vector $\vec r$ whose entries are given by $r_k=b_{\mu(k)}$ with
%\begin{widetext}
\begin{equation}
\mu(k)={s+\frac 12-\left|s+\frac 12+\sigma-k\right|}, \quad b_\mu=\frac{2^{s-1-\mu}}{1-2^s}b
\end{equation}
%\end{widetext}
for $1\le\mu\le s$, and $b_0=b$, $b_{-1}=0$, gives rise to a {\it genuine} $n$-partite
GHZ paradox for qubits.

Before the proof let us see  an example a 4-setting genuine  $6$-qubit GHZ paradox. In this case we have $s=2$, $\sigma=2$, and $b=-1/6$ and (denoting $\bar b=-b$)
\begin{equation}
\begin{array}{cccccc|c}
\hat X(0,&b,&\frac {\bar b}3,&\frac{\bar b}{6},&\frac{\bar b}{6},&\frac{\bar b}3)&+1\\
\hat X(\frac{\bar b}3,&0,&b,&\frac {\bar b}3,&\frac{\bar b}{6},&\frac{\bar b}{6})&+1\cr
\hat X(\frac{\bar b}{6},&\frac{\bar b}3,&0,&b,&\frac {\bar b}3,&\frac{\bar b}{6})&+1\cr
\hat X(\frac{\bar b}6,&\frac{\bar b}{6},&\frac{\bar b}3,&0,&b,&\frac {\bar b}3)&+1\cr
\hat X(\frac {\bar b}3,&\frac{\bar b}6,&\frac{\bar b}{6},&\frac{\bar b}3,&0,&b)&+1\cr
\hat X(b,&\frac {\bar b}3,&\frac{\bar b}6,&\frac{\bar b}{6},&\frac{\bar b}3,&0)&+1\cr
\hat X(0,&0,&0,&0,&0,&0)&+1\cr
\hat X(b,&b,&b,&b,&b,&b)&-1 \cr
\end{array}
\end{equation}

{\it Proof. --- } It is easy to check that the vector $\vec r$ defined in Theorem 2 is a GHZ vector and thus provides  a set $\mX$ of $n+\sigma$ Mermin observables with $s+\sigma$ settings via Theorem 1. Now we need to prove that when restricted to any subset $\beta$ of $3\le |\beta|<n$ qubits, no subset of $\mX$ can provide a GHZ paradox. The main idea behind our proof is that  any subset of $\mX$ does not possess a common eigenstate when restricted to a strict subset of  qubits. Suppose that this is not true and let $\alpha$ be a subset of $\mX$ and $\beta$ be a subset of qubits with $|\beta|<n$ such that the observables in $\alpha$ restricted on the qubits in $\beta$, denoted by $\hat X_\beta(\vec r)$, provide a GHZ paradox.

 For a given $s\ge\mu\ge 1$ there is always a pair of integers $k,k^\prime=n+\sigma+1-k$ satisfying $\mu(k)=\mu(k^\prime)=\mu$. As a consequence each local observable $\hat X_{b_\mu}$ of the $j$-th qubit  appears exactly twice in the collection $\mX$ of $n+\sigma$ observables, namely,  $\hat X({\vec r_{k}})$ and $\hat X({\vec r_{k^\prime}})$ with\begin{equation}\label{cons}
k^\prime=2j+\sigma+1-k \mod n:=v_j(k).
\end{equation}
That is to say for each $j\in\beta$ two observables $\hat X({\vec r_{k}})$ and $\hat X({\vec r_{v_j(k)}})$  must be either both excluded or both included in $\alpha$ with $k=0,1,\ldots,n-1$. Furthermore if $j\in\beta$ and $\hat X(\vec r)\in \alpha$ such that the local observable of  $\hat X(\vec r)$ on the $j$-th qubit is $\hat X_b$ or $\hat X_0$, then $\hat X(b\vec 1)\in\alpha$ or $\hat X\in\alpha$.

Let us consider at first the case of odd $n$, i.e., $\sigma=1$, and we shall prove at first that  $\hat X(b\vec 1)\in\alpha$. If this is not the case then for any $j\in\beta$ the local observable of any $X(\vec r)\in \alpha$ on the $j$-th qubit is not $\hat X_b$. For any given $j,l\in \beta$ since $\alpha$ is not empty there exists at least a pair of $k,k^\prime$ such that $\hat X(\vec r_k)$ and $\hat X(\vec r_{k^\prime})$ have identical local observable $\hat X_{b_{\mu_k}}$ on the $j$-th qubit. Being the unique observable that has  identical local observable on the $l$-th qubit as  $\hat X(\vec r_k)$, the observable $\hat X({\vec r_{v_l(k)}})$ is also included in $\alpha$. Furthermore the observable $\hat X(\vec r_{k_1})$ with $k_1=v_{jl}(k)=v_j\circ v_l(k)$, being the unique observable that has the identical local observable as $\hat X({\vec r_{v_l(k)}})$, should also be included in $\alpha$. We note that $v_l(k)$ is different from both $k$ and $k^\prime$ since $n$ is odd and thus $k_1$ is different from $k,k^\prime$ and also from $v_l(k^\prime)$. In the same manner $k_2=v_{jl}(k_1)$ is also different from all the integers appearing in defining $k_1$ and $v_l(k_1^\prime)$ and the observable $\hat X(\vec r_{k_2})$ should be included in $\alpha$. This process yields an infinite sequence of different observables in $\alpha$ which is impossible and therefore $\hat X(b\vec 1)\in\alpha$.

As a consequence for any $j\in \beta$ it holds $\hat X(\vec r_{j})\in\alpha$, being the unique observable whose local observable on the $j$-th qubit is exactly $\hat X_b$. Let $S_j=\{e^{i\pi\sum_{k\in\beta}\epsilon_{k}(r_{jk}-b)}\mid\epsilon_k=\pm\}$ be the spectrum of  $\hat X_\beta(\vec r_j)\hat X_\beta(b\vec 1)^\dagger$ with $r_{jk}=[[\vec r_j]]_k$. We note that $-1\not\in S_j$ since $\sum_k (|b|+|r_{jk}|)<1$  and $+1\in S_j$ if and only if i) $|\beta|$ is odd; ii) for each $j$ the components of $\vec r_j$ that are different from $b$ are paired up; iii) for each pair of paired-up components $k,k^\prime$ $\epsilon_k\epsilon_{k^\prime}=-1$. However it is impossible for all observables $\hat X_\beta(\vec r_j)\hat X_\beta(b\vec 1)^\dagger$ to have $+1$ as eigenvalue simultaneously for all $j\in \beta$. Thus the determinants of $\sum_{j\in \beta}\epsilon_j\hat X_\beta(\vec r_j)\hat X_\beta(b\vec 1)^\dagger\pm n$ are all nonzero, meaning that $\hat X_\beta(\vec r_j),\hat X_\beta(b\vec 1)$ cannot have a common eigenstate.

Let us now consider the case of even $n$. Consider the qubits set restricted  in $\beta$. If there is an observable $\hat X(\vec r)\in \alpha$ whose local observables include both $X_0$ and $X_b$, it is easy to see that $\hat X(b\vec 1)\in \alpha$ and $\pm1$ do not belong to the spectrum of $\hat X(\vec r)\hat X(b\vec 1)^\dagger$, meaning that $\hat X(b\vec 1)\in \alpha$ and $\hat X(\vec r)\in \alpha$ do not share a common eigenstate. If there is an observable $\hat X(\vec r)\in \alpha$ whose local observables include $X_b$ only then via the same argument as in the odd case we see that no common eigenstate exists for the set of observables.
If there is an observable $\hat X(\vec r)\in \alpha$ whose local observables include $X_0$ only, a similar argument excludes the possibility of a common eigenstate as the second case.
Thus we have only to consider the case where the local observables of all observables $\hat X(\vec r)\in \alpha$ are neither $\hat X_0$ nor $\hat X_b$.

Among all the pairs of integers $\{v_l(k),v_{l}({k^\prime})\}$ with $l\not=j\in\beta$ there is at most one qubit $j^\prime\in\beta$ such that $v_l(k)=v_{j^\prime}\circ v_l(k^\prime)$. This is because $k+k^\prime=2(j+1)\mod n$ has and only has two solutions, i.e., $j$ and $j+n/2$, with $k,k^\prime$ fixed in the case of even $n$.  In the case of $|\beta|=3$ this will not happen because two 3-qubit observables with two identical local observables cannot have any common eigenstate if the third local observables are not commuting. Thus there are at least one $j^\prime\in \beta$ such that $v_{jj^\prime}(k):=v_j\circ v_{j^\prime}(k)$ and $v_{jj^\prime}(k^{\prime})$ are not paired up for the $j$-th qubit and $k_1=v_j\circ v_{jj^\prime}(k)$ is different from $k,k^\prime$. By construction we have $\hat X(\vec r_{k_1})\in \alpha$. Then we start with $v_j\circ v_{jj^\prime}(k)$ and repeat the process above, which also yields an infinite sequence of different observables in $\alpha$, a contradiction. Thus we have $\hat X(b\vec 1)\in\alpha$ and, similar to the case of an odd number of qubits, we cannot find a common eigenstate of observables $\hat X(\vec r_j),\hat X(b\vec 1)$.
Therefore, it is impossible to find a new GHZ paradox by removing any qubits or removing any Mermin operators in $\mX$. Thus the GHZ paradox given by $\vec r$ as in Theorem 2 is a genuine $n$-partite
GHZ paradox. \hfill $\sharp$

As an application, we can derive an experiment-testable multi-setting Bell inequality from the multi-setting GHZ paradox.
Consider  an arbitrary GHZ vector $\vec r$ and for each $y\in Y_{\vec r}$ we introduce a two-outcome measurement $X_y$. Let us introduce a Bell operator
%\begin{widetext}
\begin{equation}\label{Bell-qubits}
  \mathcal B_b=\sum_{i=0}^{n-1} X(\vec r_i)+(\sigma-1)X-X(b\vec 1)
\end{equation}
%\end{widetext}
and in any local realistic theory it holds $\mathcal B_b\le n+\sigma-2$.  To see this we note that first, if any term in the summation in the right-hand side of Eq.(\ref{Bell-qubits}) takes the value $-1$ then the bound holds obviously. Second, if all the terms in the summation in the right-hand-side of Eq.(\ref{Bell-qubits}) take value $1$, then the last term in the right-hand side of Eq.(\ref{Bell-qubits}) must take value $-1$ and thus the bound also holds. But in quantum mechanics, the upper bound of $\mathcal B_b$ can reach the value $\langle\Gamma|\mathcal B_b|\Gamma\rangle=n+\sigma$ by choosing $|\Gamma\rangle$ to be the GHZ state $|\Phi\rangle_k$.

\section{Conclusion}

In summary, we have prescribed a general construction of GHZ paradoxes for qudit GHZ states with multisettings, i.e.,  each observer measures more than two observables while in previously known GHZ paradoxes only two alternative observables are measured by each observer. One of the advantages of introducing multisettings is that it is possible to  construct a genuine $n$-qubit GHZ paradox for any number $n\geq3$ of qubits, especially for the 4-qubit GHZ state used originally in demonstrating GHZ paradox. Also for the  experimental realizations of, e.g., $4$ qubits cases or $6$ qubits cases we have derived an  experiment-testable multi-setting Bell inequality from $2$-level GHZ paradoxes, which is maximally violated by the GHZ state. Still there is a small gap remains for 4-qudit GHZ state with each qudit of dimension $4k+2$ for $k>0$. To find  a way to construct multisetting GHZ paradoxes for other entangled states such as graph states should be of some future interests.

Note: A recent experiment was performed by a group of USTC,  and verified our framework for the 4 qubits' case. The related results will be announced soon.

\section*{Acknowledgement} 

This work is supported by National
Research Foundation and Ministry of Education, Singapore
(Grant No. WBS: R-710-000-008-271) and the NNSF of China (Grant No. 11075227 and No. 11405120) as well as the Fundamental Research Funds for the Central Universities.

\end{document}